\documentstyle[12pt,epsfig,eqsecnum,aps]{revtex}

\title{On the free volume in nuclear multifragmentation}
\author{Al. H. Raduta}
\address{National Institute of Physics and Nuclear Engineering,\\ 
  Bucharest, POB MG-6, Romania\\[.3cm]}
\begin{document}
\draft
\maketitle

\begin{abstract}
  In many statistical multifragmentation models the volume available to the
  $N$ nonoverlapping fragments forming a given partition is a basic ingredient
  serving to the simplification of the density of states formula. One
  therefore needs accurate techniques for calculating this quantity. While the
  direct Monte-Carlo procedure consisting of randomly generating the fragments
  into the freeze-out volume and counting the events with no overlapped
  fragments is numerically affordable only for partitions with small $N$, the
  present paper proposes a Metropolis - type simulation which allows accurate
  evaluations of the free volume even for cases with large $N$. This procedure
  is used for calculating the available volume for various situations. Though
  globally this quantity has an exponential dependence on $N$, variations of
  orders of magnitude for partitions with the same $N$ may be identified. A
  parametrization based on the virial approximation adjusted with a
  calibration function, describing very well the variations of the free volume
  for different partitions having the same $N$ is proposed. This
  parametrization was successfully tested within the microcanonical
  multifragmentation model from [Al. H. Raduta and Ad. R. Raduta, Phys. Rev. C
  {\bf 55}, 1344 (1997); {\it ibid.}, {\bf 56}, 2059 (1997)].  Finally, it is
  proven that parametrizations of the free volume solely dependent on $N$ are
  rather inadequate for multifragmentation studies producing important
  deviations from the exact results.
\end{abstract}

\section{Introduction}

Nuclear multifragmentation is intensely studied both theoretically and
experimentally from more than fifteen years. Since it concerns the decay of
excited systems formed in violent heavy ion collisions supposed to be
statistically equilibrated, this process has often been described by means of
statistical models (e.g. \cite{Bondorf,Gross,Randrup,FaiRan}). A basic
ingredient of some of these models is the volume disponible to $N$
nonoverlaping fragments (often called free volume and denoted herein by
$\Omega_{eff}$), positioned into a freeze-out volume $\Omega$. This serves to
the simplification of the density of states formula and implicitly to an
easier evaluation of the various statistical observables. For the case of $N$
noninteracting particles, $\Omega_{eff}$ is simply given by $\Omega^N$. When
the fragments are not allowed to overlap each other - which is the physical
case, this quantity is drastically limited. In statistical models
$\Omega_{eff}$ manifests as a factor entering the statistical weights of the
fragment partitions and thus the accuracy of the model predictions is directly
influenced by the quality of the $\Omega_{eff}$ parametrization. While an
exact analytical evaluation of the $N$-dimensional integral defining
$\Omega_{eff}$ is up to now practically impossible, various approximations for
this quantity have been reported since the beginning of the multifragmentation
studies \cite{FaiRan,Ran,Bondorf}. 

The topic of the effective volume goes far beyond the nuclear fragmentation
studies being a fundamental thermodynamical problem.  Fields like chemical
physics, solid state physics and others are dealing with such problems. Here it
is worth mentioning the approaches corresponding to infinite systems composed
of identical or non identical spherical fragments proposed more then thirty
years ago \cite{Carnahan1,Carnahan2}. While the above approaches are accurate
for the systems they have been designed for, from the point of view of the
nuclear multifragmentation studies they have two important drawbacks: they do
not account for the {\em finite} volume of the freeze-out recipient in which
fragmentation is supposed to take place and they have complicated expressions
making them unsuitable to be employed in a time consuming Metropolis sharp
microcanonical multifragmentation calculation.

So far, accurate predictions of $\Omega_{eff}$ have been performed via a
simple Monte Carlo method consisting of randomly generating the fragments into
the freeze-out volume and counting the events in which no two fragments are
overlapped \cite{Cole,DasG}. However, though it provides exact evaluations of
$\Omega_{eff}$ (within the statistical error), this method is numerically
affordable only for cases with small number of fragments. Indeed, since
$\Omega_{eff}$ is exponentially decreasing with $N$, accurate evaluations for
large $N$ are possible only with prohibitively large numbers of events.

The present paper proposes a Metropolis-type simulation procedure which makes
possible exact evaluations of $\Omega_{eff}$ even for partitions with large
number of fragments. Since variations of $\chi$ (defined as
$\Omega_{eff}/\Omega^N$) of orders of magnitude are observed for partitions
with the same $N$ but different fragments size distribution a parametrization
of $\chi$ only dependent of $N$ appears to be inadequate. For this reason a
parametrization based on an adjusted virial approximation is proposed.

The paper is organized as follows: Section \ref{sec:method} gives a
description of the standard Monte Carlo method and proposes a Metropolis type
procedure allowing exact calculations of $\chi$ even for large $N$. Section
\ref{sec:overall} presents a brief overall view on the results of the proposed
method. In Section \ref{sec:param}, a virial parametrization of $\chi$ is
proposed. Applications and tests of the resulted parametrizations within the
microcanonical multifragmentation model \cite{Noi1,Noi2} are presented in 
Section \ref{sec:application}. The conclusions are drawn in Section
\ref{sec:conclusions}.

\section{Calculating procedure}
\label{sec:method}

Suppose that we are dealing with $N$ spherical fragments of sizes
$A_i,~i=1,\dots,N$ resulted from a spherical source nucleus of size $A$:
$\sum_{i=1}^N A_i=A$, positioned at {\bf r}$_i~, i=1,\dots,N$ into a
freeze-out recipient having the volume $\Omega=n~\Omega_0$, where $\Omega_0$
is the volume of the source nucleus at normal nuclear matter density:
$\Omega_0=4 \pi r_0^3 A/3$, $r_0=1.2$ fm. The fragments are not allowed to
overlap one another or the recipient ``walls'' meaning that they are subject
to the constrains: $\prod_{i<j}^N \theta_{ij} \prod_{i=1}^N \theta^r_i$ with
$\theta_{ij}=\Theta(|{\bf r}_i-{\bf r}_j|-(R_i+R_j))$,
$\theta^r_i=\Theta(R-R_i-|{\bf r}_i|)$, where $R=r_0 A^{1/3}n^{1/3}$, $R_i=r_0
A_i^{1/3}$ and $\Theta$ is the step function. The volume available to the $N$
fragments writes:
\begin{equation}
  \Omega_{eff}(N)=\prod_{l=1}^N \int_\Omega {\rm d} {\bf r}_l~\theta^r_l
  \prod_{i<j}^N \theta_{ij}~. 
\label{eq:Omega_eff}
\end{equation}
From now on we will explicitely deal with the ratio between $\Omega_{eff}$ and
$\Omega^N$, $\chi$. As mentioned earlier, one can evaluate this quantity by
means of a simple Monte Carlo technique consisting of placing randomly each of
the $N$ fragments in the considered freeze-out recipient. The event is
considered successful if no two fragments are overlapping each other and no
fragment is intersecting the recipient walls.
Supposing that one obtains $N_s$ successful events from $N_t$ attempts, one can
estimate $\chi$ according to:
\begin{equation}
  \chi=\frac {N_s}{N_t}.
\label{eq:chi_MC}
\end{equation}
Since the relative statistical error in evaluating (\ref{eq:chi_MC}) is of the
order of $1/ \sqrt{N_s}$, it results that for accurate evaluations of small
$\chi$ values (which is the case of large $N$) one needs a very large number
of generated events ($N_t$) which makes the method unpracticable.
For this reason, a Metropolis-type simulating procedure for calculating
$\chi$ is further proposed.

Using (\ref{eq:Omega_eff}) one obtains
\begin{equation}
  \frac{\Omega_{eff}(N)}{\Omega_{eff}(N-1)} 
  =\Omega~g_N~f_N,
\label{eq:rec}
\end{equation}
and, by recurrence,
\begin{equation}
  \chi=\prod_{k=1}^N g_k~f_k,
\label{eq:chi}
\end{equation}
where
$g_k= \int_\Omega {\rm d} {\bf r}_k~\theta^r_k / \Omega
  =\left[(R-R_k)/R\right]^3$
and
\begin{equation}
  f_k=\frac{
  \int_\Omega {\rm d} {\bf r}_1~\theta^r_1
  \int_\Omega {\rm d} {\bf r}_2~\theta^r_2~\theta_{12}\dots
  \int_\Omega {\rm d} {\bf r}_k~\theta^r_k~\theta_{1k} \dots
  \theta_{k-1~k}}{\int_\Omega {\rm d} {\bf r}_1~\theta^r_1
  \int_\Omega {\rm d} {\bf r}_2~\theta^r_2~\theta_{12}\dots 
  \int_\Omega {\rm d} {\bf r}_{k-1}~\theta^r_{k-1}~\theta_{1~k-1} \dots
  \theta_{k-2~k-1}
  \int_\Omega {\rm d} {\bf r}_k~\theta^r_k}.
\label{eq:Omega_eff_1}
\end{equation} 
The last expression can be regarded from the point of view of a statistical
ensemble corresponding to a system consisting of the first $k$ ($ \in
\{2,\dots, N \}$ - we don't need to consider $k=1$ since $f_1=1$) fragments
from the considered partition, composed of configurations
$C:\{{\bf r}_i,~i=1,...,k \}$ and subject to the constrains $\prod_{i<j}^{k-1}
\theta_{ij} \prod_{i=1}^k \theta^r_i$.  This way, equation
(\ref{eq:Omega_eff_1}) can be re-written as follows:
\begin{equation}
  f_k=\frac{\sum_C \theta_{1k} \dots \theta_{k-1~k}}{\sum_C 1}
  =\left< \theta_{1k} \dots \theta_{k-1~k}\right>_k,
  \label{eq:fk}
\end{equation}
where $\left< \cdot \right>_k$ has the meaning of average value over the
states of the above mentioned ensemble.  Therefore, it is sufficient that one
knows the values of $f_k$ for $k=2,\dots,N$ in order to obtain $\chi$.  These
values can be evaluated using (\ref{eq:fk}) within a Metropolis-type
simulation which is further described.

For evaluating $f_k$ one just has to generate a Markovian walk in the
configuration space of the $k$ fragment system according to the detailed
balance principle:
\begin{equation}
\Delta C~W_C P(C \rightarrow C')=\Delta C'~W_{C'} P(C \leftarrow C'),
\label{eq:det_balance}
\end{equation}
where $\Delta C$ and $\Delta C'$ are the elementary volumes of the
configurations $C$ and $C'$, $W_C$ and $W_{C'}$ their statistical weights
(here equal to unity) and $P(C \rightarrow C')$ and $P(C \leftarrow C')$ are
the probabilities of passing from $C$ to $C'$ and respectively of making the
reverse move.

The random walk is generated as follows. Suppose that the current state of the
system is $C$. One chooses at random one of the $k$ fragments, indexed by $i$
and positioned at ${\bf r}_i$.  One randomly re-positionates this fragment into
a spherical volume having the same center as the freeze-out recipient and 
the radius $R-R_i$ (this ensures the non-overlapping between the fragment 
and the recipient wall). We denote the chosen position of the fragment $i$ with
${\bf r}'_i$. At this point one has to check whether the constrains of the 
system are satisfied. So, if $i \neq k$ and the fragment in its new position is
overlapping at least one of the remaining fragments indexed from 1 to $k-1$, 
the move is aborted and configuration $C$ is reconsidered. Otherwise, the 
configuration $C'$ is correct and is considered as a new configuration of the
system. The probability of this move is:
\begin{equation}
  P(C \rightarrow C')=\frac{{\rm d} {\bf r}'_i}{4 \pi (R-R_i)^3/3}.
\end{equation}
The probability of the reverse move, similarly generated, writes:
\begin{equation}
  P(C \leftarrow C')=\frac{{\rm d} {\bf r}_i}{4 \pi (R-R_i)^3/3}.
\end{equation}
Taking into account that $\Delta C= {\rm d} {\bf r}_i \prod_{j \neq i}^k {\rm
  d}{\bf r}_j$, $\Delta C'= {\rm d} {\bf r}'_i \prod_{j \neq i}^k {\rm d}{\bf
  r}_j$ and that $W_C=W_{C'}$, one may easily see that the detailed balance
equation, (\ref{eq:det_balance}) is satisfied.  At this point the exploration
of the configuration space is completely described. For calculating $f_k$ one
just has to apply formula (\ref{eq:fk}), i.e. to determine the average value
of the observable $ \prod_{i=1}^{k-1} \theta_{ik}$ over the states selected by
the simulation. For a number of $N_s$ {\em successful} (for which
$\prod_{i=1}^{k-1} \theta_{ik}=1$) accepted events the relative statistical
error in estimating $f_k$ is of the order of $1/\sqrt{N_s}$.

Provided that $f_k$ was evaluated for each $k\in\{2,\dots,N\}$ using the above
described simulation, $\chi$ can be obtained using (\ref{eq:chi}).  Assuming
that each simulation (corresponding to a given $k$) is performed using $N_s$
successful accepted events, then the relative statistical error in evaluating
$\chi$ is of the order of $\sqrt{N/N_s}$.

\section{Overall results}
\label{sec:overall}

\begin{figure}
  \begin{center}
  \epsfig{file=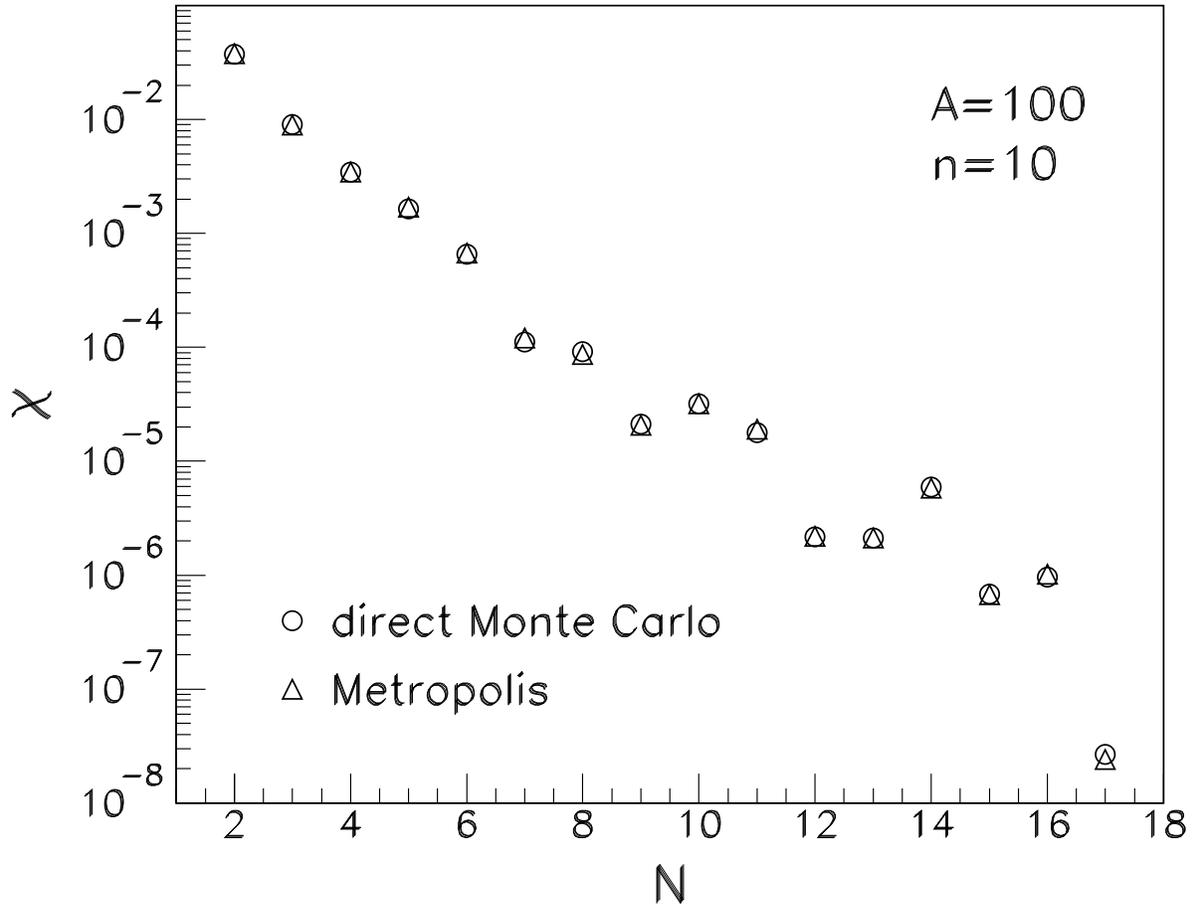,height=12cm,angle=0}
  \end{center}
  \caption{Comparison between $\chi$ values calculated by means of the 
    direct Monte Carlo procedure and those calculated by means of the
    Metropolis-type simulation for $A=100,~n=10$ and $N$ ranging from 2 to 17.
    For each $N$ a randomly generated partition was used.}
\label{fig:1}
\end{figure}

In order to verify the accuracy of the proposed simulation we present in Fig.
\ref{fig:1} a comparison between $\chi(N)$ calculated with the present
simulation and that calculated with the direct Monte Carlo procedure briefly
described at the beginning of Section I in the case of $A=100$,
$\Omega=10~\Omega_0$ (i.e. $n=10$). For this evaluation we considered $N$ in a
relatively small range (2 to 17) because for larger $N$ the direct Monte Carlo
calculation would require unreasonable computing time. The fragment partitions
corresponding to each $N$ are randomly\footnote{Starting from the source
  nucleus one randomly splits it in two fragments. Then, one randomly choses
  one of the resulted fragments and splits it randomly in two fragments. The
  process is repeated until $N$ fragments are obtained. This method has the
  advantage of generating partitions with $\chi$ taking values from a wide
  range for a given $N$.
  The dispersions of $\chi$ obtained using this generating procedure are even
  larger than those obtained using partitions realistically generated (i.e. by
  means of a microcanonical multifragmentation model). This can be easily
  checked by comparing the dispersions of $\chi$ (corresponding to $n=6$) from
  Fig. 2 with those from Fig. 8.
} selected. This calculation was performed for
$N_s=10^4$ successful events implying that the relative errors are smaller
than the dimension of the points. The same $N_s$ is used in all the
calculations presented in this paper. One can observe that the results of the
two calculations are practically identical.

\begin{figure}
  \begin{center}
  \hspace*{-1cm}  
  \epsfig{file=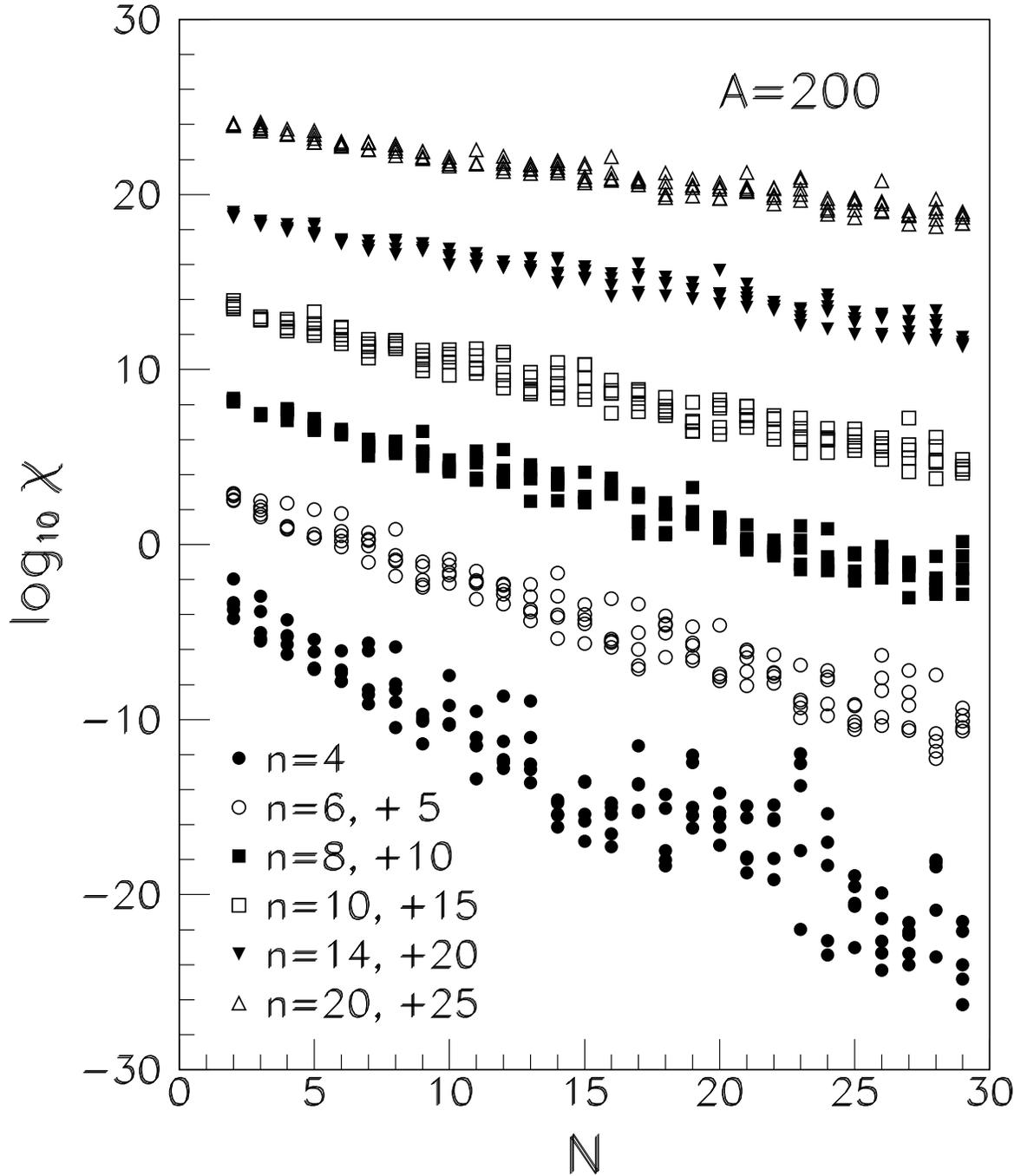,height=18cm,angle=0}
  \end{center}
  \caption{$\log_{10} \chi$  versus $N$ calculated for six values of $n$ and a 
    source with $A=200$. For each particular $n$ and $N$ 5 values of
    $\log_{10} \chi$ are represented, corresponding to randomly chosen
    partitions.  The points corresponding to $n=6,~8,~10,~14,~20$ are shifted
    upwards with the quantities specified in the legend in order to avoid
    overloading.}
\label{fig:2}
\end{figure}

For having a preliminary overall look on the simulation results, we calculate
$\chi$ for $A=200$ and $n=$4, 6, 8, 10, 14, 20 for $N$ ranging from 2 to 30
with a step of 1. For each considered $N$, 5 simulations are performed
corresponding to randomly chosen partitions. This allows to evaluate the
variation of $\chi$ for partitions having the same $N$. The results are
presented in Fig.  \ref{fig:2}. One can observe that while globally $\log_{10}
\chi$ has a linear behavior, variations of orders of magnitude, strongly
increasing with decreasing $n$ can be observed for $\chi$ corresponding to
different partitions with the same $N$.

\begin{figure}
  \begin{center}
  \hspace*{-1cm}  
  \epsfig{file=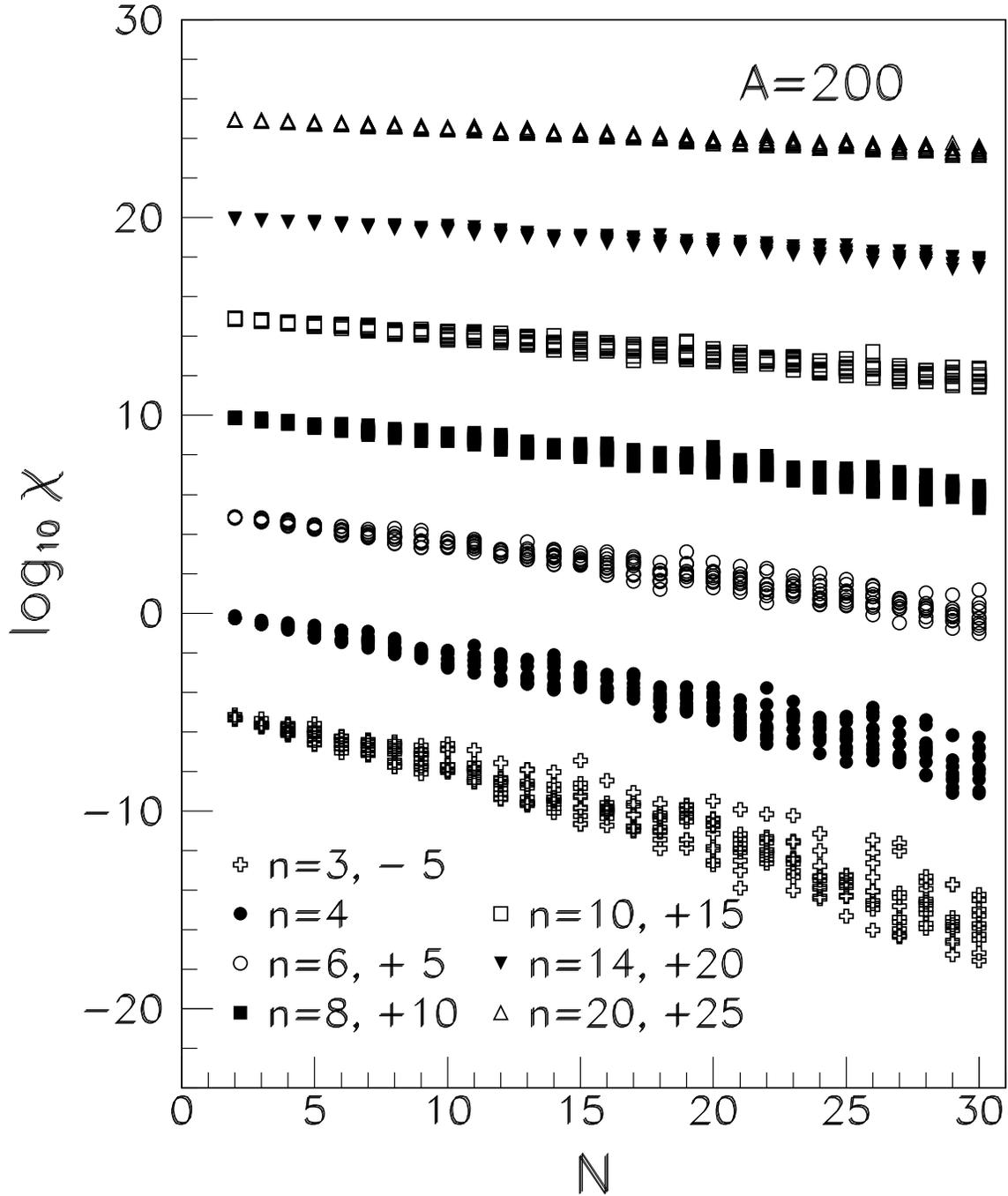,height=18cm,angle=0}
  \end{center}
  \caption{Same as in Fig. \ref{fig:2} except that the calculations are
    performed without imposing the constrain of nonintersection between
    fragments and the recipient wall. Additionally, a calculation
    corresponding to $n=3$ is included. For each particular $n$ and $N$
    10 randomly chosen partitions are used in the calculation. Points
    corresponding to various $n$ are shifted upwards or downwards as shown in
    the legend in order to avoid overloading.}
\label{fig:3}
\end{figure}

In order to test to what extent the assumption of nonoverlapping between the
fragments and the recipients' wall is contributing to the above mentioned
result, we run the same simulation without imposing this boundary any longer
(i.e. the constrains are $\prod_{i<j}^N \theta_{ij}$). Calculations of $\chi$
are performed for $A=200$ and $n=$3, 4, 6, 8, 10, 14, 20 for $N$ ranging from
2 to 30 with a step of 1 and 10 randomly chosen partitions per each $N$ value
(see \ref{fig:3}). Due to this relaxation, the effective free volume is now
larger so the dependencies $\log_{10} \chi (N)$ are less abrupt than in the
previous case (see Fig. \ref{fig:2}) and their dispersion is smaller.
Nevertheless, variations of $\chi$ of orders of magnitude corresponding to
partition with the same $N$, strongly increasing with decreasing $n$, can be
identified here as well.

This suggests that parametrizations of $\chi$
solely dependent on $N$ are quite inadequate for cases with relatively small
values of $n$ - typically used in multifragmentation studies. One therefore
needs a parametrization of $\chi$ valid over a large range of $N$, independent
of $A$ and taking into account the above-evidenced strong variation of $\chi$
when $N$ and $n$ are kept constant. This problem is addressed in the next
section.

From now on we will concentrate exclusively on the initial boundary assumption 
(i.e. nonoverlapping between the fragments and the recipient walls).

\section{Parametrization}
\label{sec:param}

The aim of this section is to provide an accurate parametrization of $\chi$
which takes into account the strong variation of this quantity for partitions
with the same number of fragments. 
Replacing the expression of $\chi$ (corresponding to the case in which the
fragments are allowed to intersect the recipients' walls):
\begin{equation}
\chi=\prod_{l=1}^N\frac{\int_{\Omega} {\rm d} {\bf r}_l \prod_{i<j}^N
  \theta_{ij}}{\int_{\Omega} {\rm d} {\bf r}_l  }
\end{equation}
by the following ``factorization'':
\begin{equation}
\chi'=\prod_{i<j}^N \frac{ \int_{\Omega} {\rm d} {\bf r}_i 
  \int_{\Omega} {\rm d} {\bf r}_j 
  ~\theta_{ij}} 
  {\int_{\Omega} {\rm d} {\bf r}_i \int_{\Omega} {\rm d} {\bf r}_j}=
  \prod_{i<j}^N \frac{ \int_{\Omega} {\rm d} {\bf r}_i 
    \int_{\Omega} {\rm d} {\bf r}_j 
  \left( 1-\bar{\theta}_{ij} \right)} 
  {\int_{\Omega} {\rm d} {\bf r}_i \int_{\Omega} {\rm d} {\bf r}_j}=
  \prod_{i<j}^N \left( 1-P_{ij} \right),
\label{eq:two-body1}
\end{equation}
where $\bar{\theta}_{ij}=1-\theta_{ij}$ and $P_{ij}=(\int_{\Omega} {\rm d}
{\bf r}_i \int_{\Omega} {\rm d} {\bf r}_j~\bar{\theta}_{ij})/ (\int_{\Omega}
{\rm d} {\bf r}_i \int_{\Omega} {\rm d} {\bf r}_j)$, one easily recognizes the
two-body approximation due to Cole, Heuer and Charvet \cite{Cole}. Note that
here $P_{ij}$ has the meaning of probability of overlapping between the
particles $i$ and $j$ when they are randomly generated into the 
volume $\Omega$. A more convenient form for (\ref{eq:two-body1}) is:
\begin{equation}
  \ln \chi'=\sum_{i<j}^N \ln \left( 1-P_{ij} \right)
\end{equation}
which for small values of $P_{ij}$ can be approximated by:
\begin{equation}
  \ln \chi_v= - \sum_{i<j}^N P_{ij}.
  \label{eq:virial}
\end{equation}
The last expression is the so called virial approximation proposed by Randrup,
Robinson and Sneppen \cite{Ran}. While the exact expression of the two
particle overlapping probability, $P_{ij}$, was deduced in \cite{Cole}, we
prefer the simplest approach \cite{Ran}:
\begin{equation}
P_{ij}=\left( \frac{R_i+R_j}R \right)^3.
\end{equation}
Obviously, since it neglects the higher order interactions between fragments
and allows intersections of the fragments with the recipients' walls, this
approximation overestimates $\chi$. 
Nevertheless, apart from other approaches (see e.g. \cite{Carnahan1,Carnahan2}
used in chemical physics calculations) the ``virial'' approximation has the
advantage that it is very simple and that it explicitely accounts for the
volume of the freeze-out recipient (thus being appropriate for
multifragmentation studies). As we shall further see, this approach provides
the basis for a very accurate parametrization of the free volume.

\begin{figure}
  \begin{center}
  \hspace*{-.7cm}  
  \vspace*{1cm}  
  \epsfig{file=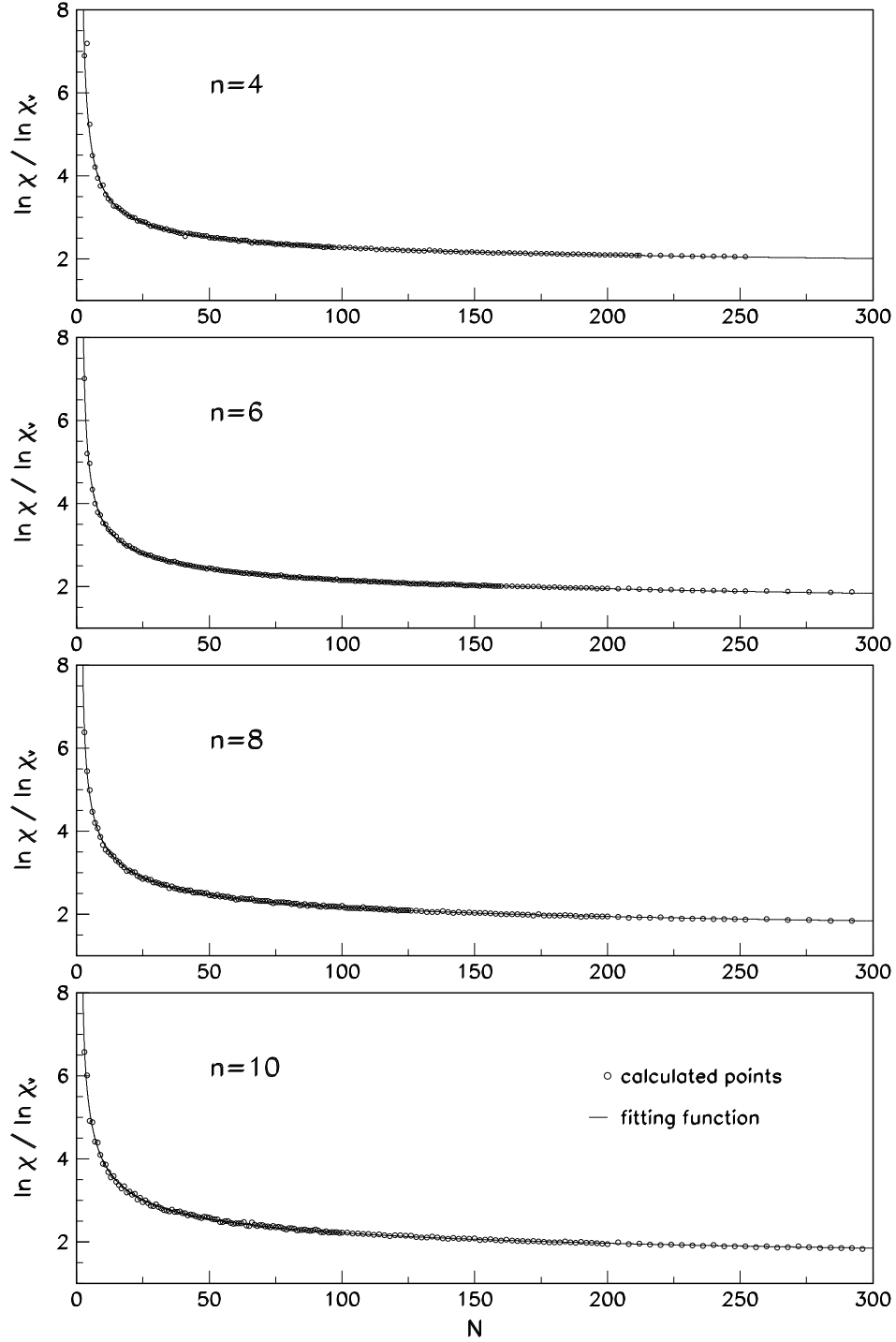,height=19cm,angle=0}
  \end{center}
  \caption{Ratio between $\ln \chi$ and $\ln \chi_v$ (see the text) 
    versus the number of fragments ($N$) calculated for $A=300$ and
    $n=4,~6,~8,~10$ fitted with $f(N)$ (see equation (\ref{eq:f(x)})). The
    corresponding fitting parameters are listed in Table \ref{table:1}.}
\label{fig:4}
\end{figure}

\begin{figure}
  \begin{center}
  \hspace*{-.7cm}  
  \vspace*{2cm}  
  \epsfig{file=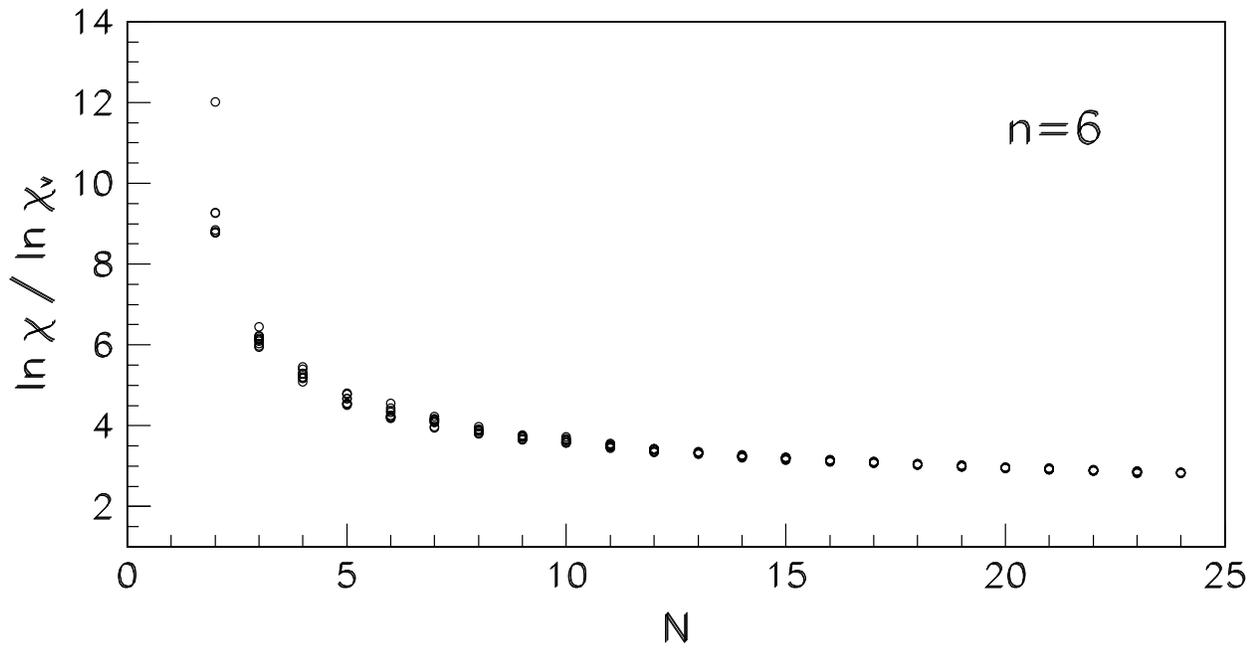,height=8.5cm,angle=0}
  \end{center}
  \vspace*{-1.5cm}
  \caption{$\ln \chi / \ln \chi_v (N)$ calculated for $A=300$, $n=6$ and 10
    randomly chosen fragment partitions for each value of $N$.}
\label{fig:5}
\end{figure}

In order to estimate the difference between the virial evaluations ($\chi_v$)
and the exact $\chi$, evaluated by means of the method proposed in Section
\ref{sec:method}, we represented in Fig. \ref{fig:4} the ratio 
$\ln \chi / \ln \chi_v$
calculated over a large range of $N$ (sufficient for the systems currently
considered in multifragmentation studies), for four values of $n$: 4, 6, 8 and
10. For each considered $N$ a fragment partition was generated as in Section
\ref{sec:overall}. The considered source is $A=300$. Though the fragment
partitions are generated as to induce an important dispersion in $\chi$ (see
Fig. \ref{fig:2}), $\ln \chi / \ln \chi_v$ has a smooth behavior. 
In order to have an estimate of the fluctuations of the $\ln \chi / \ln \chi_v
(N)$ dependency, we evaluate this quantity for $n=6$, $A=300$, $N$ ranging
from 2 to 24 and 10 randomly chosen partitions for each value of $N$. (see
Fig. \ref{fig:5}).  Though the fluctuations are increasing with decreasing $N$
they lay in reasonable limits and are practically negligible for $N>10$.

\begin{figure}
  \begin{center}
  \hspace*{-.7cm}  
  \vspace*{2cm}  
  \epsfig{file=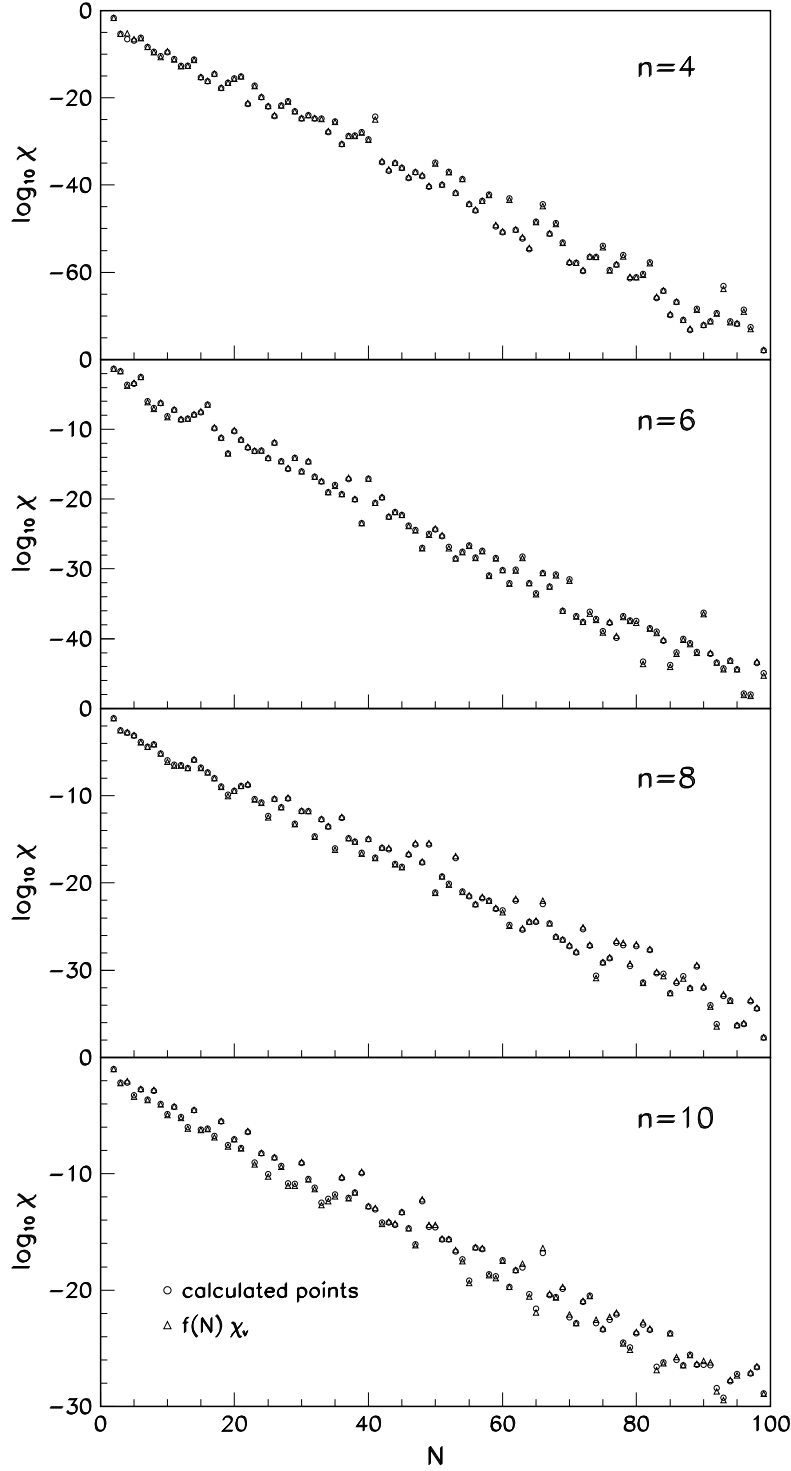,height=19.5cm,angle=0}
  \end{center}
  \vspace*{-1.5cm}
  \caption{Calculated $\log_{10} \chi$ versus $N$  corresponding to $A=300$
    and $n=4,~6,~8,~10$. For each $N$, $\chi$ was calculated for a randomly
    (see Section \ref{sec:overall}) chosen partition. The circles correspond
    to calculations using the exact method from Section \ref{sec:method}
    and the triangles correspond to the adjusted virial approximation
    [eq. (\ref{eq:param})] using the parameters from Table \ref{table:1}.}
\label{fig:6}
\end{figure}

Therefore, one can obtain accurate fits of $\ln \chi / \ln \chi_v (N)$
with appropriate
functions. As shown in Fig. \ref{fig:4}, the function:
\begin{equation}
  f(N)=\frac{a~N+b}{N^c+d~N^e}
  \label{eq:f(x)}
\end{equation}
provides a good fit for $\ln \chi / \ln \chi_v$ versus $N$, for all considered
$n$. The corresponding parameters are listed in Table \ref{table:1}. Knowing
$f(N)$ for a given $n$, the real $\ln \chi$ can be expressed as:
\begin{equation}
  \ln \chi=\ln \chi_v~f(N),
  \label{eq:param}
\end{equation}
with $\ln \chi_v$ given by (\ref{eq:virial}). In order to test this
parametrization, in Fig. \ref{fig:6} are represented versus $N$ $\chi$
calculated using the method from Section II and $\chi$ evaluated by means of
(\ref{eq:param}) for $n=$4, 6, 8, 10. The fragment partitions were
considered as in the previous paragraph.  The range of $N$ is considered
only up to $N=100$ just for the clarity of the plot. As one can see, the
agreement is remarkably good. It thus appears that the present parametrization
has the ability of describing accurately the strong variations of $\chi$ for
different partitions having the same $N$. Moreover, since for a given
partition $\chi$ only depends on the fragment {\em relative} size
distribution, it results that this parametrization is independent on $A$. Some
applications and tests of this parametrization are presented in the next
section.

\section{Application to a microcanonical multifragmentation model}
\label{sec:application}

In this section some supplemental tests for the accuracy of the
parametrization proposed in Section \ref{sec:param} performed with the sharp
microcanonical multifragmentation model from Ref.  \cite{Noi1,Noi2} are 
presented.
This model sharply conserves the number of nucleons ($A$) the number of
protons ($Z$), the total energy ($E$) and the total momentum (${\bf P}$) and
assumes equal probability of appearance of all possible configurations
$C_1:\{A_i,~Z_i,~{\bf r}_i,~\epsilon_i,~{\bf p}_i,~~i=1,\dots,N \}$, where the
parameters under the brackets are respectively the mass number, the atomic
number, the position, the excitation energy and the momentum of each of the
$N$ fragments composing the respective configuration. Integrating the
expression of the total number of states of the system over the fragment
momenta one reduces the configuration space and works into a new one, composed
of configurations $C:\{A_i,~Z_i,~{\bf r}_i,~\epsilon_i,~~i=1,\dots,N \}$
having statistical weights of the form:
\begin{equation}
  W_C=\frac1{N!}\prod\limits_{i=1}^N
  \left[\frac{\rho_i(\epsilon_i)}{h^3}
    \left(mA_i\right)^{3/2}
  \right]
  \frac{2\pi}{\Gamma(\frac 32(N-1))} 
  \frac{(2\pi K)^{\frac 32N-\frac 52}}{(mA)^{\frac 32}},
\end{equation}
where $\rho_i$ is the level density of the fragment $i$, $m$ is the mass of a
nucleon, $\Gamma()$ is the hyper-geometric function and $K$ is the kinetic
energy of the given partition and can be expressed as: $K=E+\sum_i B_i -
\sum_i \epsilon_i - \sum_{i<j} V_{ij}$. Here $B_i$ is the binding energy of
the fragment $i$ and $V_{ij}$ is the two-body Coulomb interaction energy. The
weight of a configuration $C':\{A_i,~Z_i,~\epsilon_i,~~i=1,\dots,N \}$
corresponding to a new space where the fragment positions are now missing, can
be written as:
\begin{equation}
  W_{C'}=\prod_{l=1}^N \int_{\Omega} {\rm d}{\bf r}_l~W_C~
  \theta^r_l \prod_{i<j}^N \theta_{ij},
  \label{eq:new_w:1}
\end{equation} 
where $\Omega$, $\theta^r_l$ and $\theta_{ij}$ have the same meaning as in the
previous sections. Now, it can be observed that if one replaces the multi-body
Coulomb interaction depending on the fragment positions (entering in $K$ and
therefore in $W_C$) by a single particle mean-field approach, $W_C$ 
nolonger depends on ${\bf r}_i$ and (\ref{eq:new_w:1}) can be rewritten as:
\begin{equation}
  W_{C'}=\Omega^N~\chi~W_C.
  \label{eq:new_w:2}
\end{equation} 
As one can see, in this case $\chi$ manifests as a factor entering the
statistical weight of a given configuration, $C'$. The average value of any
system observable $X$ is evaluable in this new ensemble by: $\left< X
\right>=\sum_{C'} W_{C'} X_{C'} / \sum_{C'} W_{C'}$. For doing this one applies
identically the Metropolis simulation proposed in \cite{Noi1,Noi2} except that the
positions are no longer generated. The resulted correction factor, $\alpha$ is
the same as in Ref. \cite{Noi1,Noi2} except for the factor 
$\chi(C'_{N+1}) / \chi(C'_N)$ appearing now due to
the modification of the configuration weights (see equation. (\ref{eq:new_w:2})).
For the Coulomb interaction the Wigner-Seitz approach is  employed adjusted
with a factor as to provide accurate descriptions of the total Coulomb energy
even in cases with small $N$:
\begin{equation}
  V_C=V_{WS}~g(N),
\end{equation}
with
\begin{equation}
  V_{WS}=\frac35 \ \frac{Z^2 e^2}{R}-\sum\limits_i \frac35 \ 
  \frac{Z_i^2 e^2}{R_i^c},
\end{equation}
where $R_i^c=R_i~n^{1/3}$ , $V_{WS}$ is the Wigner-Seitz energy and the factor
$g(N)=(1.07675~N+18.5266)/(N+14.9439)$ correspond to $n=6$, which is the case
for which the microcanonical calculations are here performed. This factor was
evaluated by fitting the ratio between the average Coulomb energy (evaluated
by uniformly generating the fragments from a given partition into the spherical
recipient such as they are not overlapping each other - this is
performed by simple Metropolis moves) and the corresponding $V_{WS}$ versus
$N$. It is worth mentioning that, included in the microcanonical model, this
parametrization insures practically identical results with those obtained
with the unmodified version of the model.

For testing the accuracy of the free-volume parametrization two types of
simulations are performed via the microcanonical multifragmentation model.
In the first simulation we just replace the multi-body Coulomb interaction
energy with the Wigner-Seitz approach exposed above. Except this detail, the
simulation is identical with that from Refs. \cite{Noi1,Noi2} and therefore the
fragment positions are explicitely generated at each step, the forbidden
configurations (in which fragments are overlapping) being automatically
rejected. The second simulation is that formulated in the previous
paragraph and includes both the Wigner-Seitz parametrization and the
free-volume parametrization. In principle, a correct parametrization of $\chi$
should provide closely identical results for any system observable. Here we
choose to compare the mass distributions calculated within the two
simulations for the source nucleus (70, 32) and the excitation 
energies $E_{ex}$=2 MeV/nucleon and $E_{ex}$=6 MeV/nucleon.
These are represented in Fig. \ref{fig:6}. As one can see, the agreement is
very good for both cases fact which supports the present parametrization. 

\begin{figure}
  \begin{center}
  \epsfig{file=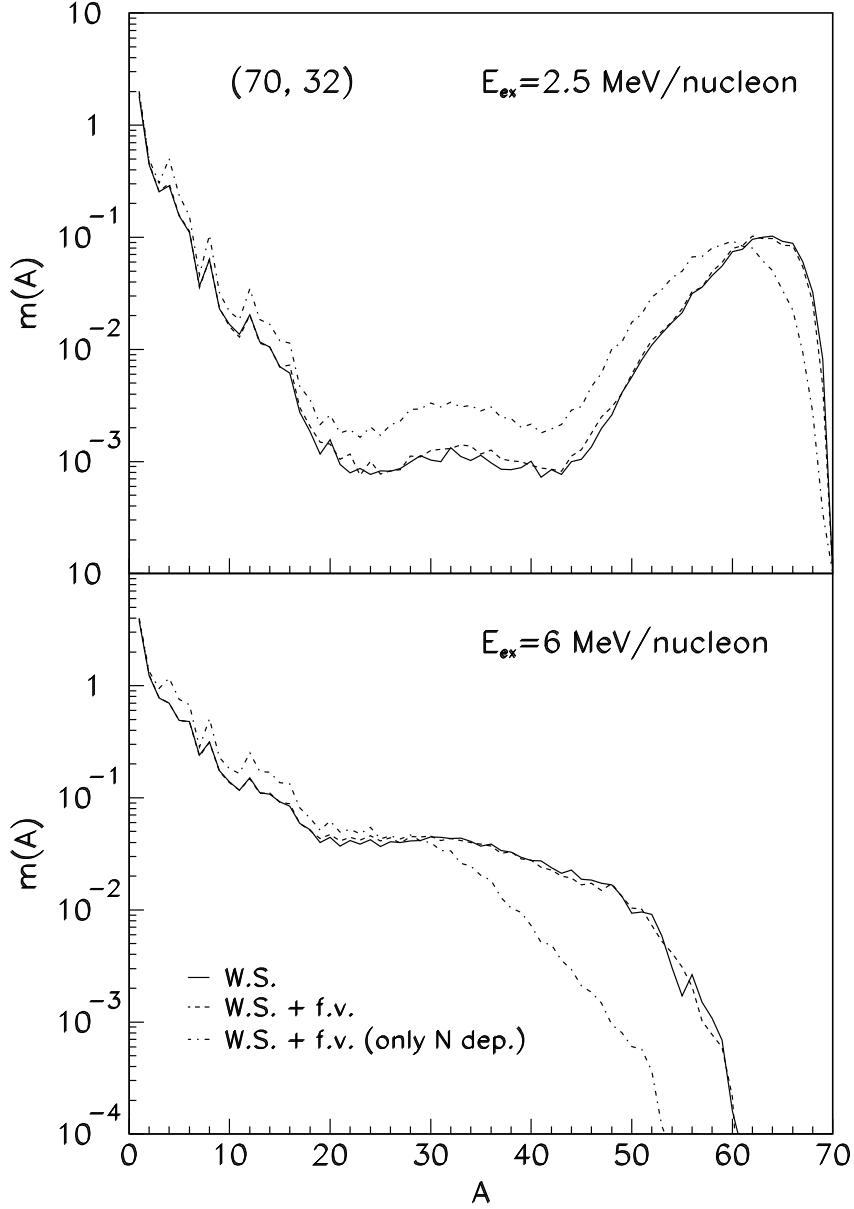,height=16cm,angle=0}
  \end{center}
  \caption{Fragment mass multiplicities  corresponding
    to the source (70,32) and the excitation energies: 2.5 and 6 MeV/nucleon
    calculated using the microcanonical multifragmentation model. The
    continuous lines are calculated using the adjusted Wigner-Seitz (W.S.)
    approach (see the text) but keeping the fragment positioning into the
    freeze-out volume, the dashed lines are calculated using both the W.S.
    approach and the free volume (f.v.) parametrization (\ref{eq:param}) and
    the dashed-dotted lines correspond to the W.S. approach and a f.v.
    parametrization only dependent on $N$ (see Fig. \ref{fig:6}).}
\label{fig:7}
\end{figure}

\begin{figure}
  \begin{center}
  \epsfig{file=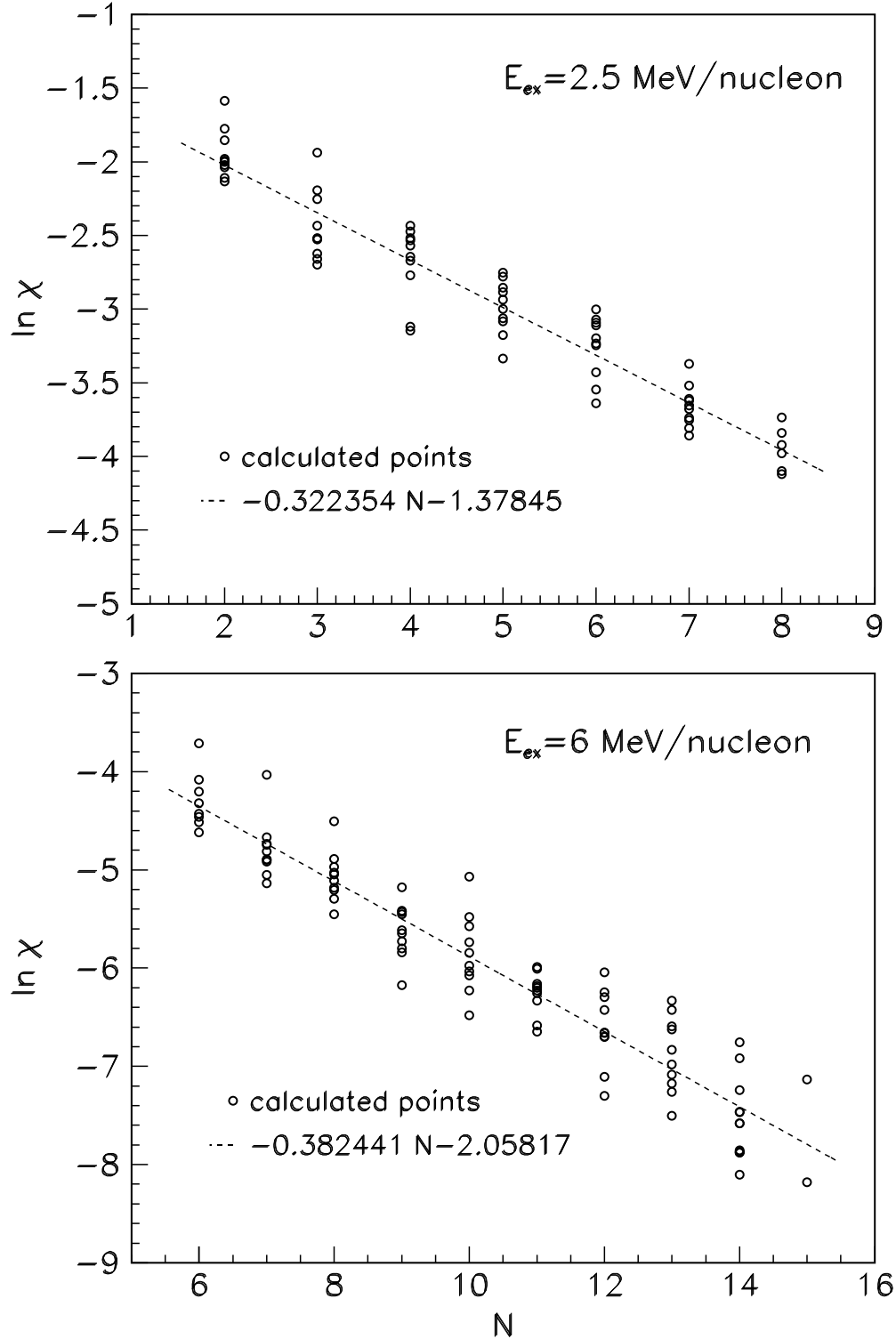,height=20cm,angle=0}
  \end{center}
  \caption{$\ln \chi$ versus $N$ calculated for the fragment partitions
    appearing in the fragmentation of the (70,32) source nucleus for the
    excitation energies 2.5 and 6 MeV/nucleon. Linear fits of $\ln \chi (N)$
    are represented by dashed lines. The partitions corresponding to each $N$
    are randomly selected from the output of the microcanonical model in which
    the W.S. approach was used.}
\label{fig:8}
\end{figure}

An interesting question is to what extent a free-volume parametrization only
dependent on $N$ is appropriate for nuclear multifragmentation studies. For
answering this question one needs to obtain for each of the above considered
cases a parametrization of $\chi$ only dependent on $N$. To this aim, within
the first of the two simulations mentioned in the previous paragraph, for each
situation [(70, 32), $E_{ex}$=2,~6 MeV/nucleon] we select at random 
different fragment partitions, corresponding to each
$N$ appearing in the simulation, for which we
calculate $\chi$.  These are represented in Fig. \ref{fig:7}. Since the
obtained dependence is rather linear in logarithmic scale we fit these
distributions by means of linear functions (see Fig. \ref{fig:7}). We then
employ the obtained parametrizations $\chi(N)$ in the multifragmentation model
and we perform the second type of simulation described in the previous
paragraph. The results are represented in Fig. \ref{fig:7} by dashed-dotted
lines. As one can observe, the difference between the mass distributions
obtained with this last simulation and the mass distribution corresponding to
the first simulation (mentioned in the previous paragraph) are quite
significant. This confirms the fact that the dispersion in $\chi$ for
different partitions having the same $N$ which was illustrated in Figs.
\ref{fig:2}, \ref{fig:8} produces indeed important deviations of the model
results from the real values and, therefore, a parametrization of $\chi$
solely dependent on $N$ appears to be inadequate. This result is important
since it is well known that most of the existent multifragmentation models
employ parametrizations of $\chi$ only dependent of $N$.

\begin{figure}
  \begin{center}
  \epsfig{file=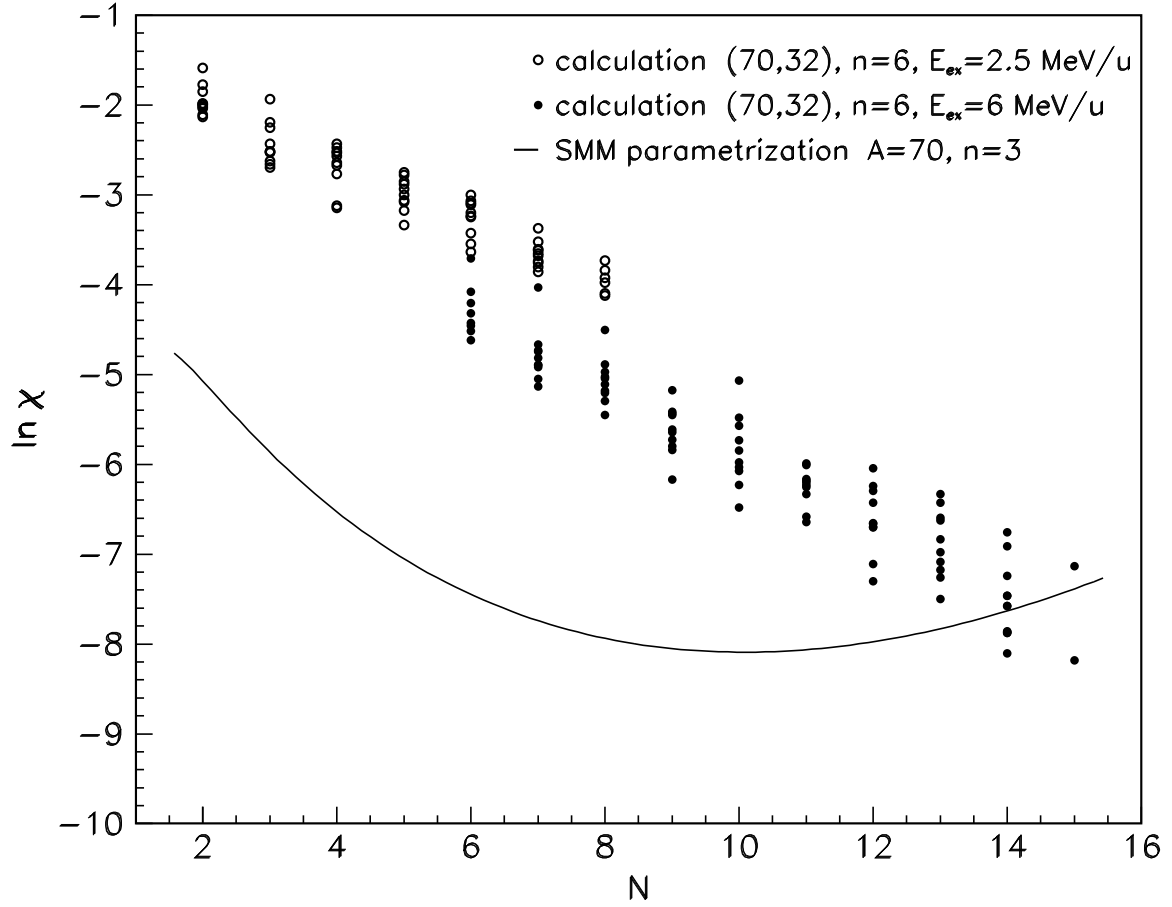,height=12cm,angle=0}
  \end{center}
  \caption{$\ln \chi (N)$ calculated for the fragment partitions
    appearing in the fragmentation of the (70,32) source nucleus for the
    excitation energies 2.5 and 6 MeV/nucleon and $\ln \chi (N)$ corresponding
    to the SMM free volume parametrization for the case of $A=70$ and $n=3$.}
\label{fig:9}
\end{figure}

Finally, a comparison between $\chi (N)$ corresponding to the partitions
considered in the previous paragraph (for the source (70, 32) and
$E_{ex}$=2,~6 MeV/nucleon) and the SMM parametrization \cite{Bondorf} is
performed. According to the SMM prescription one has: $\chi=\{
[1+d(N^{1/3}-1)/(r_0 A^{1/3})]^3-1 \} ^N / n^N$ where d=1.4 fm.  The $\chi
(N)$ dependencies calculated as described in the previous paragraph are
represented in Fig. \ref{fig:9} together with the SMM one calculated for $n=3$
- the usual SMM assumption (since in the SMM model the free volume
parametrization has no explicit dependence on $n$ we prefer to use its 
standard freeze-out assumption, $n=3$). Analyzing Fig. \ref{fig:9} one can 
observe that the SMM $n=3$ curve is flatter than
our calculations performed for the $n=6$ case. Taking into account that the
$\ln \chi(N)$ slope is increasing with decreasing $n$ (see Fig. \ref{fig:2}),
it follows that the differences between the SMM curve and the exact results
corresponding to $n=3$ are even larger than those observed in Fig. \ref{fig:7}
between the exact calculations with $n=6$ and the SMM $n=3$ curve.

\section{Conclusions}
\label{sec:conclusions}

In summary, a Metropolis-type procedure was designed in order to describe the
volume available to $N$ spherical fragments positioned into a spherical
recipient. Apart from the direct Monte Carlo technique, this method allows one
(for the first time)
to make $\chi$ evaluations even for large number of fragments. Calculations of
$\chi$ for randomly chosen partitions corresponding to various $n$ and $N$
values are evidencing that while, for a given $n$, $\chi$ has a more or less
global linear dependence of $N$ in logarithmic scale, variations of orders of
magnitude may be identified for $\chi$ calculated for different partitions
corresponding to the same $n$ and $N$. This suggests that parametrizations of
$\chi$ only dependent on $N$ may be inadequate for multifragmentation studies.
For this reason a parametrization based on virial approximation \cite{Ran}
adjusted with a calibration function as to fit the exact $\chi$ values is
proposed further-on. The obtained parametrization describes the $\chi$
variations corresponding to different partitions having the same $N$
remarkably well. Calibration parameters have been evaluated for four values of
the volume of the recipient, corresponding to $n=4,~6,~8,~10$. The
parametrization is further tested with the microcanonical multifragmentation
model from Ref. \cite{Noi1,Noi2}. For doing this, the simulation is run in two
forms: first with the inclusion of the Wigner-Seitz approach for the Coulomb
interaction but maintaining the fragment positioning into the freeze-out
volume and the hard core repulsion and then with including both the
Wigner-Seitz approach and the free-volume parametrization obtained in Section
\ref{sec:param}. Mass distributions obtained with these two types of
simulation are compared in two input situations.  A very good agreement is to
be noticed which supports once again the parametrization obtained in Section
\ref{sec:param}. Finally a study is made concerning the extent to which
a parametrization of $\chi$ only dependent on $N$ is appropriate for
multifragemtation studies. To this aim, the second simulation was run again
this time using a parametrization of $\chi$ only dependent on $N$. In this
last case, important deviations between the mass yields obtained with the
first simulation and those obtained with the second one are to be noticed.
This suggests parametrizations of $\chi$ only dependent on $N$ may induce
important deviations of the models' results from the real ones.
Comparisons between the evaluated $\chi(N)$ dependencies and the corresponding
SMM parametrization show important discrepancies.

Presently, there is a strong debate concerning the role of the freeze-out
volume in determining the behavior of various thermodynamical quantities
(such as the nuclear caloric curve). It was shown that the inclusion of a free
volume parametrization in a statistical multifragmentation model is crucial
for a good theoretical description of various experimental isotopic caloric
curves \cite{Gulminelli}. The value of the freeze-out volume was proven to
influence the shape of the caloric curve and even the order of a possible
phase transition: while larger volumes are generating caloric curves with
continuously increasing plateau-like regions \cite{Noi2} (reflected in
positive peaks in the heat capacity curves), smaller freeze-out volumes seem
to encourage the occurrence of back-bendings in the caloric curves, reflected
in negative regions of the heat capacity curves \cite{Noi3}.  Even the
definition of the freeze-out volume is still intensely discussed: should the
statistical ensemble be isochore (see e.g. \cite{Gross}) or should it be isobar
(which implies a smooth variation of the freeze-out volume with the excitation
energy - as shown in Ref. \cite{Bondorf-rc} this option leads to the
appearance of back-bendings in the caloric curves) - see
\cite{Bondorf}. Recently, nonspherical shapes of the freeze-out recipient 
have been investigated \cite{LeFevre} as well. 
Recent experimental evaluations of the freeze-out volume using two particles 
velocity correlations  predicted values of $n$ in a wide 
range: 2.5 - 12.5 \cite{Aladin}.

\smallskip

\noindent
The author thanks Ad. R. Raduta for a critical read of the manuscript.
\newpage

\begin{table}
  \vspace*{2cm}
\begin{tabular}{cccccccc}
$n$&$a$       & $b$       & $c$      & $d$       & $e$      \\
\hline
4  & 2.189191 &  1.096606 & 1.042162 & -0.934725 & 0.711251 \\
6  & 3.665597 & -0.599464 & 1.125311 & -1.202542 & 0.432149 \\
8  & 0.927222 & -1.519126 & 0.999927 & -1.085739 & 0.863371 \\
10 & 1.117942 & -1.976695 & 1.020913 & -1.100722 & 0.870260 \\
\end{tabular}
\vspace*{.3cm}
\caption{The parameters of the function $f(N)$ [(\ref{eq:f(x)})] 
  corresponding to four values of $n$.}
\label{table:1}
\end{table}

\end{document}